\documentclass{pasa}

\usepackage{definitions}
\usepackage{amsmath,amsbsy}
\usepackage{graphicx}

\usepackage{subfigure}

\title[Bias-free model fitting of correlated data in interferometry]{Bias-free model fitting of correlated data in interferometry}
\author[Lachaume]{R\'egis Lachaume$^{1,2}$\thanks{regis.lachaume@gmail.com}
\affil{$^{1}$Instituto de Astronom\'\i{}a, Facultad de F\'\i{}sica, Pontificia Universidad Cat\'olica de Chile, casilla 306, Santiago 22, Chile}%
\affil{$^{2}$Max-Planck-Institut f\"ur Astronomie, K\"onigstuhl 17, D-69117 Heidelberg, Germany}%
}

\jid{PASA}
\doi{TBD}
\jyear{\the\year}
\usepackage{aas_macros}

\usepackage{hyperref}
\hypersetup{colorlinks,citecolor=blue,linkcolor=blue,urlcolor=blue}

\begin{document}

\begin{frontmatter}
\maketitle

\begin{abstract}
In optical and infrared long-baseline interferometry, data often display significant correlated errors because of uncertain multiplicative factors such as the instrumental transfer function or the pixel-to-visibility matrix.  In the context of model fitting, this situation often leads to a significant bias in the model parameters. In the most severe cases this can can result in a fit lying outside of the range of measurement values. This is known in nuclear physics as Peelle's Pertinent Puzzle.  I show how this arises in the context of interferometry and determine that the relative bias is of the order of the square root of the correlated component of the relative uncertainty times the number of measurements. It impacts preferentially large data sets, such as those obtained in medium to high spectral resolution.  I then give a conceptually simple and computationally cheap way to avoid the issue: model the data without covariances, estimate the covariance matrix by error propagation using the modelled data instead of the actual data, and perform the model fitting using the covariance matrix. I also show that a more imprecise but also unbiased result can be obtained from ignoring correlations in the model fitting.
\end{abstract}

\begin{keywords}
techniques: interferometric --- 
methods: statistical ---
methods: data analysis
\end{keywords}

\end{frontmatter}

\section{Introduction}

Optical and infrared long-baseline interferometry consists in measuring the fringe contrast and phase of interference fringes in the recombined light collected at several telescopes\footnote{Recombination may be performed by software in the case of intensity interferometry or heterodyne detection.}. These observables hold information on the celestial object's spatial properties, often obtained through model fitting. 

In spite of strong evidence of correlations in the data, due to redundancy \citep[][in the case of closure phases]{MON07}, calibration \citep{PER03}, or atmospheric biases acting on all spectral channels in the same way \citep{LAW00}, only a few authors \citep{PER04,ABS06,BER06,LAC19,KAM20} have accounted for these correlations while most assumed statistically independent errors.  In particular, the only interferometric instrument I know of with a data processing software taking into account one source of correlations---calibration---is FLUOR\footnote{Fiber Linked Unit for Optical Recombination} \citep[at IOTA\footnote{Infrared and Optical Telescope Array}, then CHARA\footnote{Center for High Angular Resolution Array}, ][]{PER04}. None of the five first and second-generation ones at the VLTI\footnote{Very Large Telescope Interforometer} does \citep{AMBER,MIDI,PIONIER,GRAVITYpipe,MATISSEpipe}.  The same lack of support for correlations is present in image reconstruction programmes \citep[e.g. MIRA, see][]{THI08}, model-fitting tools \citep[e.g. Litpro, see][]{TAL08}, or the still widespread first version of the Optical Interferometric FITS format \citep[OIFITS v. 1,][]{OIFITS1}.

Unfortunately, ignoring correlations may lead to significant errors in model parameters as \citet{LAC19} evidenced with stellar diameters using PIONIER\footnote{Precision Integrated-Optics Near-infrared Imaging ExpeRiment} \citep{PIONIER} data at the VLTI. Also \citet{KAM20} established that accounting for correlations is necessary to achieve a higher contrast ratio in companion detection using GRAVITY \citep{GRAVITY} at the VLTI. 

Several sources of correlated uncertainties occur in a multiplicative context, when several data points are normalised with the transfer function \citep{PER03} or the coherent fluxes are derived with the pixel-to-visibility matrix formalism \citep{TAT07}.  In both cases, the uncertainty on the multiplicative factor translates into a systematic, correlated one in the final data product. In the context of experimental nuclear physics, \citet{PEE87} noted that this scenario could lead to an estimate falling below the individual data points, a paradox known as Peelle's Pertinent Puzzle (PPP). It results from the usual, but actually incorrect, way to propagate covariances, in which the measured values are used in the calculations \citep{DAG94,NEU12}.  A few workarounds have been proposed but they are either computationally expensive \citep[e.g. sampling of the posterior probability distribution for Bayesian analysis, see][]{NEU12} or require a conceptually difficult implementation \citep{BEC12,NIS14}.  

The issue, however, is not widely known in many other fields where the problem has seldom arisen.  In this paper, I present the paradox within the context of long-baseline interferometry (Sect.~\ref{sec:ppp}), derive the order of magnitude of its effect using the modelling of a single value (Sect.~\ref{sec:single}), analyse in detail its effect in least squares model-fitting (Sect.~\ref{sec:model}) and propose a simple, computer-efficient way to avoid it (Sect.~\ref{sec:conclusion}).  

\section{Peelle's Pertinent Puzzle}
\label{sec:ppp}
I rewrite and adapt Peelle's original example in the context of long-baseline interferometry \citep[see][Sect.~1 \& 2.1]{NEU12}. One or several calibrator observations yield the inverse of the instrumental fringe contrast $\cot \pm \cot\reldev$.  I use the relative uncertainty $\reldev$ on the transfer function as it is often referred to in percentage terms. A visibility amplitude is now estimated from two contrast measurements $\raw_1 \pm \absdev$ and $\raw_2 \pm \absdev$.  For each measurement, the visibility amplitudes are:

\begin{subequations}
\begin{align}
    \data_1 &= \system{\cot\raw_1}{\cot\absdev}{\cot\raw_1\reldev},\\
    \data_2 &= \system{\cot\raw_2}{\cot\absdev}{\cot\raw_2\reldev},
\end{align}
\label{eq:data12}
\end{subequations}
where the second-order error term $\cot\reldev\absdev$ has been ignored.

They are normalised with the same quantity (\cot), so they are correlated, hence the systematic uncertainty term between parentheses in \eqref{data12}. Error propagation yields the covariance matrix

\begin{figure}
\includegraphics[width=\linewidth]{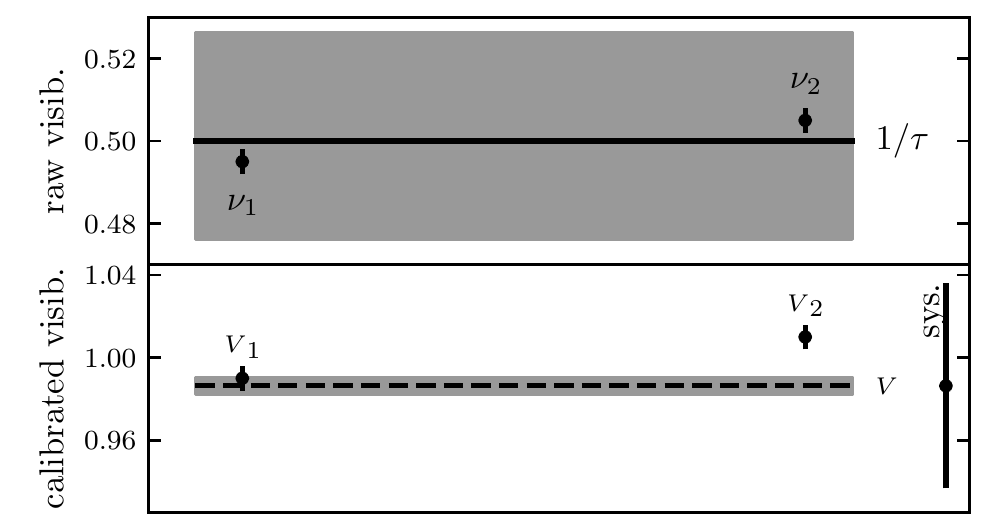}
\caption{Original Peelle problem rewritten in the context of interferometry.  \emph{Top:} Two raw visibility amplitudes $\nu_1$ and $\nu_2$ (points with statistic error bars of $\approx 0.6\%$) are calibrated by the transfer function $1/\tau$ (solid line with systematic error zone of 5\%). \emph{Bottom:} the two calibrated visibility measurements $\data_1$ and $\data_2$ (points with statistic error bars of $\approx 0.6\%$) are strongly correlated. The least-squares estimate for the visibility $\data$ (dashed line, with statistic uncertainty error zone displayed) falls outside of the data range.  The systematic error on $\data$, $\data_1$, and $\data_2$ is shown on the right.}
\label{fig:peelle}
\end{figure}

\begin{equation} 
   \vcov = \begin{pmatrix} 
     \absdev^2\cot^2 + \reldev^2\data_1^2 & \reldev^2\data_1\data_2\\
     \reldev^2\data_1\data_2              & \absdev^2\cot^2 + \reldev^2\data_2^2
            \end{pmatrix}.
\end{equation}
Under the hypothesis of Gaussian errors, I obtain the least squares estimate using the weight matrix $\vec{W} = \vcov^{-1}$:
\begin{gather}
\begin{split}
    {\data} &= 
                \frac{  \data_1              W_{11} 
                      + (\data_1 + \data_2)  W_{12} 
                      + \data_2              W_{22} 
                     }
                    {   W_{11} 
                      + 2 W_{12} 
                      + W_{22}  
                    },
                 \\
            &= \frac{\data_1 + \data_2}{2} 
            \left(1 + \reldev^2\frac{(\data_1-\data_2)^2}{2\tau^2\absdev^2}\right)^{-1},
\end{split} \label{eq:mu2}\\
\intertext{with the uncertainty}
\begin{split}
    \datadev^2 &= \frac{1}{W_{11} + 2W_{12} 
                     + W_{22}},\\
               &=   \left[ 
                        \left(\frac{\tau\absdev}{\sqrt{2}}\right)^2 
                      +  \frac{\data_1^2 + \data_2^2}{2}\reldev^2 
                    \right]
                    \frac{2\data}{\data_1 + \data_2}. \label{eq:datadev}
\end{split} 
\end{gather}
The visibility estimate $\data$ systematically falls below the average of the two values $\data_1$ and $\data_2$.  If the measurements differ significantly, it can even fall below the lowest value.  Figure~\ref{fig:peelle} gives such an example with an instrumental visibility of 50\% and two measurements on an unresolved target: 
\begin{align*}
    \cot   &= 2.000 \pm 0.100\ (\reldev = 5\%),\\
    \raw_1 &= 0.495 \pm 0.003\ (\absdev/\raw_1 \approx 0.6\%),\\
    \raw_2 &= 0.505 \pm 0.003\ (\absdev/\raw_2 \approx 0.6\%),
\intertext{which yields two points 2.4 standard deviations apart}
    \data_1 &= \system{0.990}{0.006}{0.050},\\
    \data_2 &= \system{1.010}{0.006}{0.051}
\intertext{and the visibility amplitude estimate}
    \data &= \system{0.986}{0.004}{0.050}
\end{align*}
falls outside the data range.  The uncertainties quoted for $\data$ correspond to the first and second terms within the square brackets of Eq.~(\ref{eq:datadev}).

\begin{table}
\caption{Symbols used in this paper. Lower case bold font is used for vectors and upper case bold font for matrices.}
\begin{tabular}{ll}
\hline\hline
Symbol                        & Meaning\\
\hline
$\mean a$                     & true value of $a$\\
$<a>$                         & expected value of $a$\\
$\tr{\vec A}$                 & transpose of $\vec A$\\
$\vec c = \vec a\hadam\vec b$ & element-wise product of $\vec a$ and $\vec b$\\
$\vec C = \vec a\outer\vec b$ & outer product of $\vec a$ and $\vec b$\\
$\vec c = \vec A\vec b$       & matrix product of $\vec A$ and $\vec b$\\
$\delta$                      & Kronecker delta\\
\hline
$\vdata$                      & data\\ 
$\verror$                     & error ($= \vdata - \vdatamean$)\\
$\rerror$                     & relative error ($= \error / \vdatamean$)\\
$\dev$                        & deviation ($= \sqrt{\expect{\error^2}}$), uncertainty\\
$\rdev$                       & relative uncertainty ($= \sqrt{\expect{\rerror^2}}$)\\
$\vcov$                       & covariance matrix ($= \expect{\verror\outer\verror}$)\\
$\corr$                       & correlation coefficient\\
$\vsens$, $\msens$            & sensitivity vector or matrix\\
$\vparam$                     & parameters of the model\\
$\vmod$                       & model values ($= \msens\vparam\approx\vdatamean$)\\
\hline
$a\meas$                      & $a$ of the measurement error\\
$a\norm$                      & $a$ of the normalisation error\\
$\ppp a$                      & $a$ impacted by PPP\\
$\sysdev$                     & relative systematic uncertainty\\
$\stadev$                     & relative statistical uncertainty\\
\hline
\end{tabular}
\end{table}

\section{Fit by a constant}
\label{sec:single} 

\begin{figure}
\centering
\includegraphics[width=\linewidth]{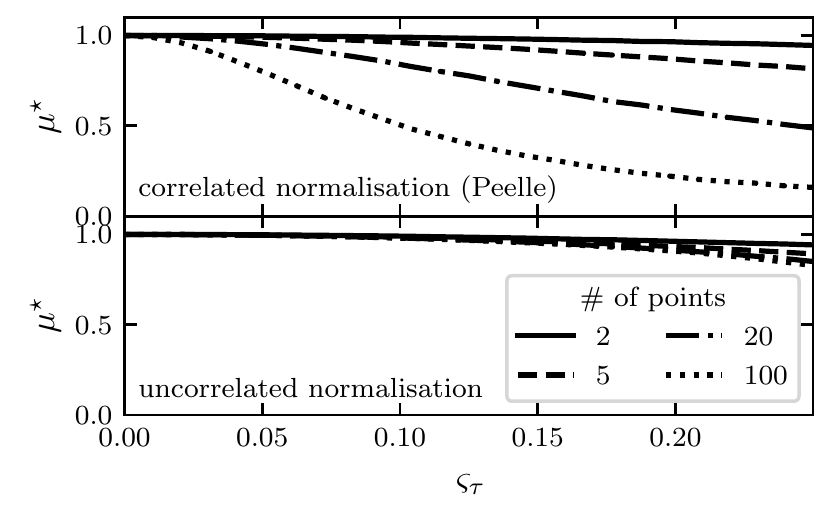}
\caption{Fit $\datappp$  to unresolved visibilities ($\data = 1$), as a function of the relative uncertainty on the calibration $\reldev$ and the number of measurements $n$. $2/n\times10^5$ simulations were made and averaged, assuming that $\rawerror$ and $\relerror$ follow normal distributions. \textit{Top:} fully correlated normalisation like in original Peelle's puzzle ($\rawdev=0.02$ and $\corr = 1$). \textit{Bottom:} normalisation error without correlation ($\rawdev=0.02$ and $\corr = 0$).}
\label{fig:uncorr-peelle}
\end{figure}

I now generalise the results of the last section to an arbitrary number of measurements of a single normalised quantity, such as the visibility amplitude of an unresolved source, which is expected to be constantly one for all interferometric baselines. Let the column vector $\vdata = \tr{(\data_1, \cdots, \data_n)}$ contain the $n$ visibility amplitudes. It is derived from an uncalibrated quantity like the fringe contrast, $\vraw = \tr{(\raw_1, \cdots, \raw_n)}$ and a normalisation factor, like the cotransfer function, $\vcot = \tr{(\cot_1, \cdots, \cot_n)}$ by 
\begin{equation}
    \vdata = \vcot\hadam\vraw,
\end{equation}
where $\hadam$ denotes the Hadamard (element-wise) product of vectors. With $\datamean$, $\cotmean$, and $\rawmean$ the true, but unknown, values of these quantities, the error vector on $\vdata$  
\begin{align}
    \verror    &= \vdata - \datamean, \\
\intertext{can be written as a sum of measurement and normalisation relative errors if one ignores the second-order terms:}
    \verror    &= (\vrawerror + \vrelerror) \datamean.\\
\intertext{These errors are given by}
    \vrawerror &= \frac1\rawmean (\vraw - \rawmean), \\
    \vrelerror &= \frac1\cotmean (\vcot-\cotmean).
\end{align}

I assume $\vrawerror$ and $\vrelerror$ are independent, of mean 0, and have standard deviations $\rawdev$ and $\reldev$, respectively. In addition, I consider correlation of the normalisation errors, with correlation coefficient $\corr$.  In the case of interferometry, it can arise from the uncertainty on the calibrators' geometry.  The covariance matrix of the visibility amplitudes is given by
\begin{align}
    \vcov     &= \expect{\verror\outer\verror},\\
\intertext{where $\outer$ denotes the outer product of vectors and $\expect{}$ stands for the expected value, so that}
    \cov_{ij} &= \underbrace{\left[ \rawdev^2 
                  + (1-\corr)\reldev^2 \right]}_{\stadev^2} \datamean^2
                \delta_{ij} 
            + \underbrace{\corr\reldev^2}_{\sysdev^2} \datamean^2.
\end{align}
The non-diagonal diagonal elements of the matrix feature the systematic
relative uncertainty $\sysdev$, i.e. the correlated component of the
uncertainties.  In the case of a fully correlated transfer function ($\corr =
1$), it is equal its uncertainty ($\sysdev = \reldev$). The diagonal term of
the matrix additionally includes the statistical relative uncertainty
$\stadev$, i.e. the uncorrelated component of the uncertainties.  In the case
of a fully correlated transfer function, it is equal to the uncertainty of the
uncalibrated visibility ($\stadev = \rawdev$).

The value $\datamean$ is yet to be determined, so the covariances are often derived using the measurements $\vdata$ in the propagation:
\begin{equation}
    \ppp{\cov_{ij}} = \stadev^2 \data_i^2 \delta_{ij} 
            + \sysdev^2\data_i\data_j. \label{eq:cov}
\end{equation}

The least squares estimate for $\datamean$ is given by
\begin{equation}
    \ppp{\mod} =
                  \frac { \tr\vsens{\ppp\vcov}^{-1}\vdata }
                        { \tr\vsens{\ppp\vcov}^{-1}\vsens }
    \label{eq:pppmod}
\end{equation}
where $\vsens = \tr{(1, \cdots, 1)}$ is the trivial sensitivity vector.  The covariance matrix is the sum of an invertible diagonal matrix and one of rank one---see \eqref{cov}---, so that the inverse is obtained using the Woodbury matrix identity: 
\begin{align}
    \invcov{ij} &= \frac{\delta_{ij}}{\dev_i^2}
         - \frac{\sysdev^2\data_i\data_j}
                {\dev_i^2\dev_j^2\Big(1 + 
            \sysdev^2\sum\limits_{k} \frac{\data_k^2}{\dev_k^2}\Big)},
            \label{eq:invcov}\\
  \intertext{where we have introduced the statistical (uncorrelated) component of the uncertainty on the calibrated visibilities}
  \dev_i     &= \stadev\data_i \label{eq:devi}.
\end{align}

Appendix~\ref{ap:mu} shows the analytical derivation for the least squares estimate $\ppp{\mod}$ using the previous formulae. I write it in a way that highlights the generalisation of Eq.~(\ref{eq:mu2}) of the previous section:
\begin{equation}
  \ppp{\mod} = 
    \frac{\sum\limits_{i} \frac{\data_i}{\dev_i^2}}{\sum\limits_i \frac{1}{\dev_i^2}}
    \left( 1 + \sysdev^2\,\frac
            {\sum\limits_{i<j} \frac{(\data_i-\data_j)^2}{\dev_i^2\dev_j^2}} 
            {\sum\limits_i \frac{1}{\dev_i^2}}
        \right)^{-1}.
    \label{eq:mu}
\end{equation}

For small enough errors ($\error_i \ll \datamean$) the second-order Taylor development in $\error_i = \data_i - \datamean$ yields (see Appendix~\ref{ap:mutaylor}): 
\begin{equation}
    \ppp{\mod} \approx 
        \datamean
      + \frac1n {\sum\limits_i \error_i}
      - \frac 1\datamean 
        \left(\frac1{2n}\frac{\sysdev^2}{\stadev^2}  + \frac 1{n^2}\right) 
      \sum\limits_{i\ne j} (\error_i-\error_j)^2.
    \label{eq:mutaylor}
\end{equation}

Since $\expect{\error_i} = 0$ and $\expect{(\error_i-\error_j)^2} = 2\stadev^2\datamean^2$, the expected value
\begin{equation}
    \expect{\ppp{\mod}}  \approx \datamean 
            \left[ 1 -  
               \left(1-\frac1n\right)
               \left(2\stadev^2 + n\sysdev^2\right)
               \right]
    \label{eq:mutaylor2}
\end{equation}
is biased.  If the data are not correlated ($\sysdev = 0$), the bias is small
($\stadev^2$ to $2\stadev^2$) but it becomes larger for correlated data if the number of points is large ($n/2$ to $n \times \reldev^2$ for fully correlated data) as \citet{DAG94} already noted. This analytical derivation confirms the numerical simulation by \citet[][see their Fig.~4]{NEU12}. For visualisation purposes, Figure~\ref{fig:uncorr-peelle} shows a similar simulation of the bias as a function of the normalisation uncertainty $\reldev$ for various data sizes ($n = 2$ to 100). I have verified that it reproduces the quadratic behaviour of Eq.~(\ref{eq:mutaylor2}) for small values of $\stadev$ and $\sysdev$ (bias inferior to 10 to 20\% of \datamean).

The bias from PPP arises, intuitively, because the modelled uncertainty is a non-constant function of the measured value. In the present case, data that fall below the average are given a lower uncertainty and, thus, a higher weight in the least squares fit. Conversely, data that fall above the average have a higher uncertainty and a lower weight. This fundamentally biases the estimate towards lower values.  The effect is much stronger with correlations because it impacts the $\sim n^2/2$ independent elements of the covariance matrix instead of being restricted to the $n$ diagonal ones.  In the literature, the puzzle is generally discussed as arising from a normalisation, as it it where it has been first identified. However, I show in appendix~\ref{ap:photon} that it is not necessary and determine the bias in the case of correlated photon noise.

For spectro-interferometric observations with 4 telescopes, the number of correlated points can be over 1,000, so even with a low correlation coefficient, the bias can be significant. For instance, a single GRAVITY observation in medium spectral resolution yields $n = 6 \times 210$ visibility amplitudes. With an observed correlation of $\corr \approx 16$\% in the instrumental visibility amplitudes \citep{KAM20} and a typical $\reldev = 1$--2\% normalisation error, the bias on the calibrated visibilities could be 2--8\%.  

\section{General model fitting}
\label{sec:model}

\begin{figure}
\centering
\subfigure[Under-resolved, linear least squares]{\includegraphics[width=\linewidth]{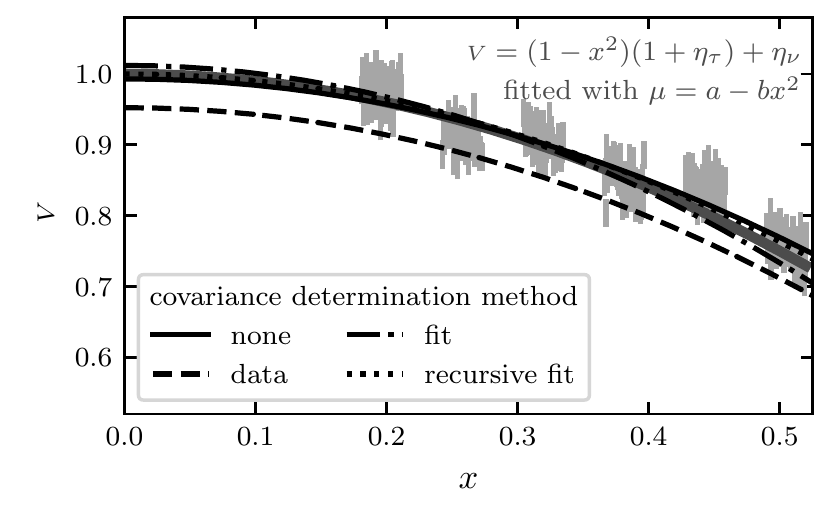}\label{fig:fitexample}}\\
\subfigure[Well resolved, non-linear least squares]{\includegraphics[width=\linewidth]{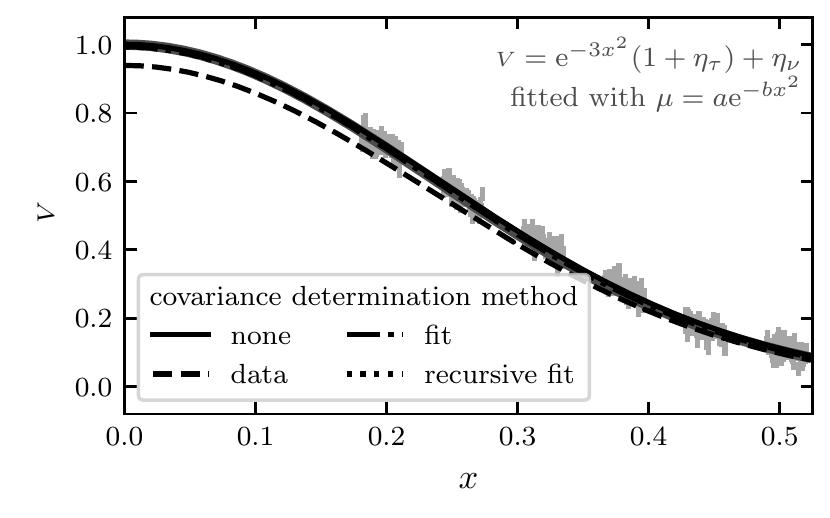}\label{fig:fitexample2}}
\caption{Model fitting to simulated correlated data from a four-telescope interferometer (6 baselines) with medium spectral resolution (R = 100) with 2\% uncorrelated measurement error and 3\% correlated normalisation error (light gray points with the measurement error bar).  \emph{Top:} Simulated under-resolved data $\data = 1-x^2$ (thick gray line) are fitted with linear least-squares model $\mod = a-bx^2$ using the four prescriptions for the covariance matrix. \emph{Bottom:} The same for well-resolved data $\data = \exp -3x^2$ and non-linear least-squares with model $\mod = a\exp -bx^2$.}
\end{figure}

\begin{figure*}
\centering
\subfigure[Under-resolved data fitted with a linear least squares model]{\includegraphics[width=.8\linewidth]{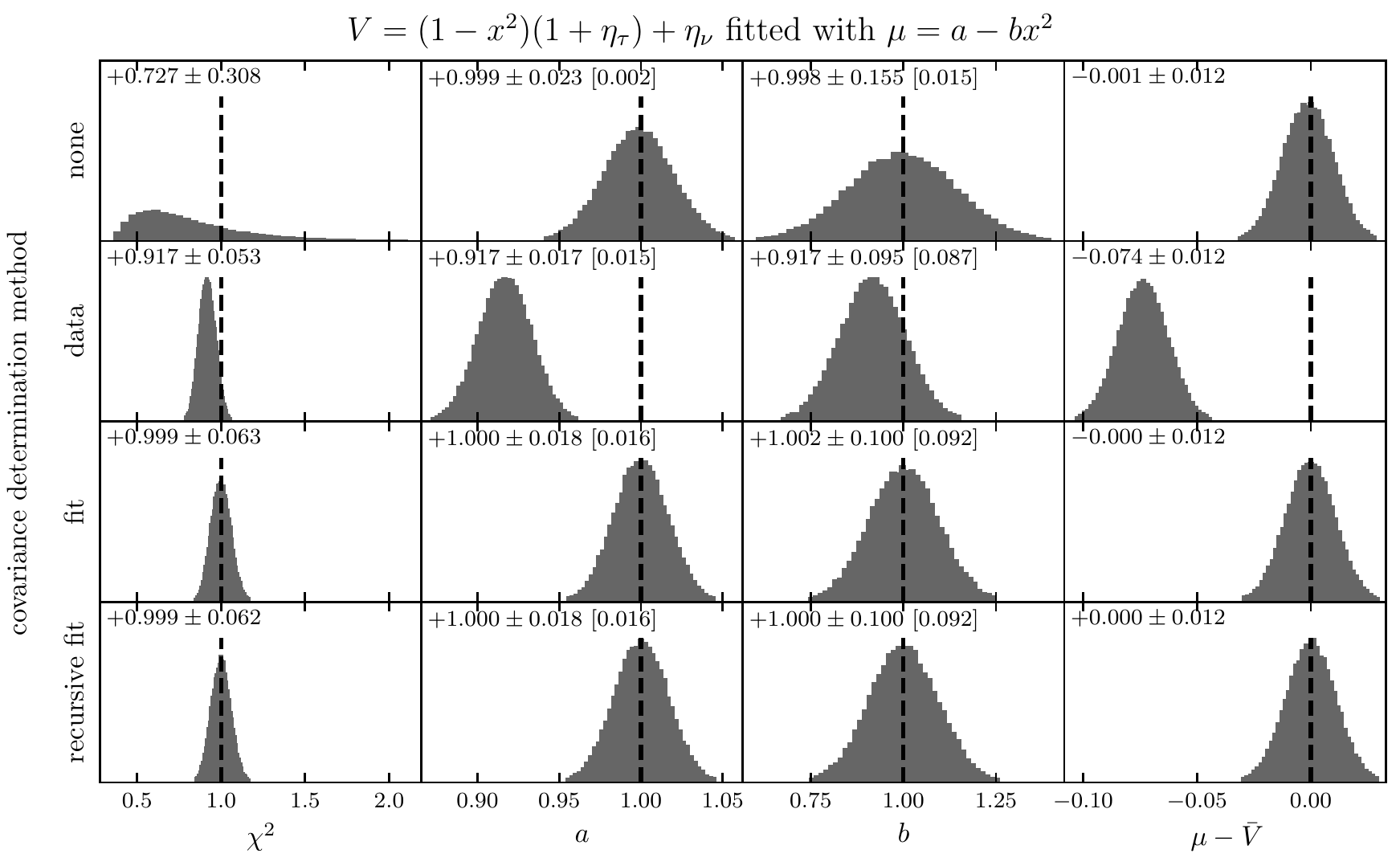}\label{fig:fitanalysis}}\\
\subfigure[Well-resolved data fitted with a non-linear least squares model]{\includegraphics[width=.8\linewidth]{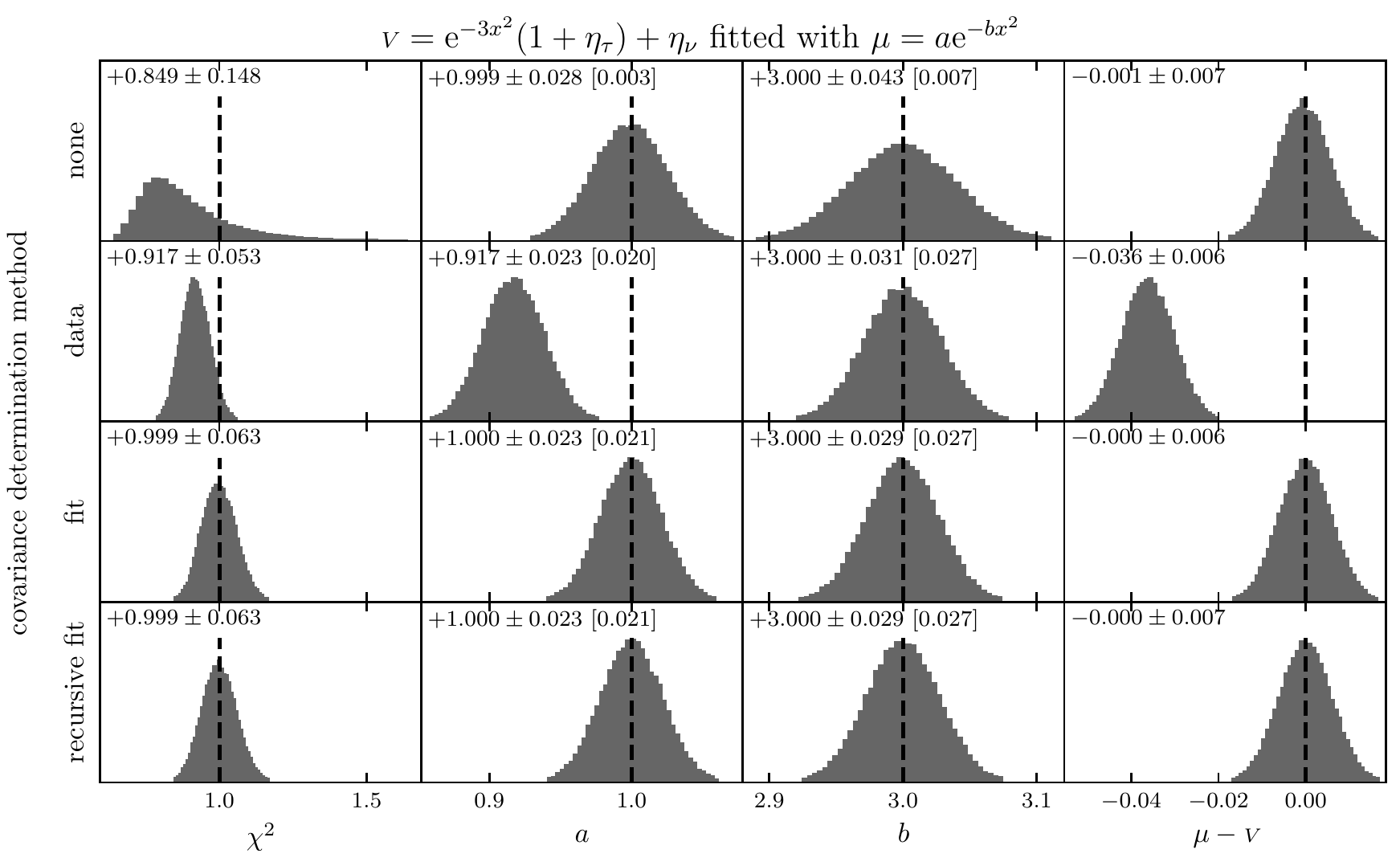}\label{fig:fitanalysis2}}
\caption{Distribution of the fitted parameters and fit properties for the four covariance matrix prescriptions analysed in Sect.~\ref{sec:model}. $5\times10^4$ simulations of 6 groups of 100 correlated data points $\data$ (measurement error $\rawdev = 2\%$ and normalisation error $\reldev = 3\%$, correlation of the latter $\corr = 1$, normal distributions) are performed and fitted with a model using least-squares minimisation. \emph{Top graph:} under-resolved data follow $\data = 1-x^2$ and are fitted with linear least squares $\mu = a - bx^2$. \emph{Bottom graph:} well resolved data follow $\data = \exp -3x^2$ and are fitted with non-linear least squares $\mu = a\exp -bx^2$. Reported quantities include median and 1-$\sigma$ interval of their distribution and, within brackets, the median uncertainty reported by the least squares fit. 
The covariance matrix prescriptions  are: \emph{top row:} correlations are ignored; \emph{second row:} a na\"\i{}ve covariance matrix uses the data values; \emph{third row:} covariance matrix uses modelled values from fit without correlations; \emph{bottom row:} covariance matrix and model are recursively computed, with the covariance matrix of the next recursion using the modelled value of the last step. }
\end{figure*}

I now consider a set of measurements corresponding to the linear model 

\begin{align}
    \vmod  &= \msens\vparam,\\
\intertext{where $\vparam$ are the unknown parameters and $\msens$ is the known sensitivity matrix. Typically, $\sens_{ik}=f_k(u_i, v_i)$ for a linear model and $\sens_{ik} = \partial f/\partial p_k(u_i, v_i)$ for a non-linear model approximated by a linear one close to a solution. $(u, v)$ is the reduced baseline projected onty the sky, i.e. $u = B_u / \lambda$ if $\vec B$ is the baseline and $\lambda$, the wavelength.  The true values $\vdatamean$ are impacted by errors so that the data are} 
    \vdata &= \vdatamean + \verror\\
\intertext{with the error term $\verror$ again expressed as the sum of a measurement and a normalisation error:}
    \verror &=  \vabserror + \vrelerror \hadam \vdatamean.
\end{align}
The measurement errors $\vec\abserror$ and normalisation errors $\vec\relerror$ follow  multivariate distributions of mean zero with covariance matrices $\vabscov$ and $\vrelcov$ respectively. Given the covariance matrix $\vcov$ of this model, the least squares estimate is
\begin{align}
   \vparam &= (\tr\msens \vcov^{-1} \msens)^{-1} (\tr\msens \vcov^{-1} \vdata)
\end{align}
I investigate four ways to determine the covariance matrix $\vcov$
\begin{enumerate}
    \item Ignoring the correlations in the normalisation using $\vcov_0 = \vabscov + (\vdata\outer\vdata)\hadam(\vrelcov\hadam\vec I)$. Let $\vdataflat = \msens(\tr\msens\vcov_0^{-1}\msens)^{-1}\tr\msens\vcov_0^{-1}\vdata$ the resulting model of the data.\label{item:nocorr} 
    \item Using the na\"\i{}ve estimate $\ppp\vcov = \vabscov + (\vdata\outer\vdata)\hadam\vrelcov$ which is known to lead to Peelle's pertinent puzzle in the trivial case of a constant model.\label{item:ppp}
    \item Using the data model of the fit without the normalisation error: $\vcov_1 = \vabscov + (\vdataflat\outer\vdataflat)\hadam\vrelcov$. This is the generalisation of the two-variables approach by \citet{NEU14}. The resulting least squares model is $\vdatarec{1} = \msens(\tr\msens\vcov_1^{-1}\msens)^{-1}\tr\msens\vcov_1^{-1}\vdata$.\label{item:fit}
    \item Recursively fitting the data by updating the data model in the covariance matrix. I derive $\vdatarec{k} = \msens(\tr\msens\vcov_k^{-1}\msens)^{-1}\tr\msens\vcov_k^{-1}\vdata$ using $\vcov_{k} = \vabscov + (\vdatarec{k-1}\outer\vdatarec{k-1})\hadam\vrelcov$, starting with the estimate $\vdatarec{1}$ ($k = 2$).\label{item:recfit}
\end{enumerate}

In order to compare these covariance matrix prescriptions, I will use the typical example of an under-resolved centro-symmetric source observed at a four-telescope facility in medium spectral resolution. It is close to the context under which I serendipitiously noticed the effect while modelling stellar diameters \citep[see][]{LAC19}. The \textsc{python} code to produce the results (figures in this paper) is available on github.\footnote{\url{https://github.com/loqueelvientoajuarez/peelles-pertinent-puzzle}} In the under-resolved case all models---Gaussian, uniform disc, or limb-darkened disc---are equivalent \citep{LAC03}, so I will use instead a linear least squares fit $\mod = a - bx^2$ to $\data \approx 1 - x^2$ where $x$ is dimensionless variable proportional to the projected baseline length $\sqrt{u^2+v^2}$. This fit corresponds to the second-order Taylor development of any of the aforementioned models.  Figure~\ref{fig:fitexample} shows the example of such a fit performed for each covariance matrix prescription.  Data have been simulated using $\data = (1-x^2)(1 + \relerror) + \abserror$ where $\relerror$ is a fully correlated normalisation error (3\%) and $\abserror$ are uncorrelated statistical errors (2\%). As expected, the use of data $\vdata$ in the correlation matrix, method \ref{item:ppp}, leads to grossly underestimated data values, in the very same way as in the classical Peelle case described in Sects.~\ref{sec:ppp}~\&~\ref{sec:single}.  Other methods, including ignoring correlations, yield reasonable parameter estimates.

Figure~\ref{fig:fitanalysis} sums up the behaviour of the same fit performed a large number of times on different simulated data sets, each following $\data \approx 1-x^2$.  For each correlation matrix prescription, it displays the dispersion of the reduced chi squared, the model parameters $a$ and $b$, and the difference between modelled value and true value. It also reports the uncertainty on model parameters given by the least squares optimisation routine in comparison to the scatter of the distribution of the values. While the model fitting ignoring correlations (method~\ref{item:nocorr}) does not show any bias on the parameter estimates, it displays a higher dispersion of model parameters,  grossly underestimates the uncertainty on model parameters and has a biased chi squared.  The correlation matrix calculated from data (method~\ref{item:ppp}) is, as expected, strongly biased. Both methods estimating the correlation  matrix from modelled data (methods~\ref{item:fit}~\& \ref{item:recfit}), are equivalent in terms of the absence of bias, dispersion of these quantities, and correct prediction of the uncertainty on model parameters.

Given that fitting recursively the covariance matrix doesn't yield additional benefits for the modelling, I would suggest to use method \ref{item:fit}. One would expect this to hold for any smooth enough model, as the update in the covariance matrix is expected to be a small effect. Indeed, I have checked that the result holds for a fully resolved Gaussian disc $\data \approx \exp -3x^2$ fit by $\mod = a\exp -bx^2$ (see Figs.~\ref{fig:fitexample2}~\&~\ref{fig:fitanalysis2}) \and a well-resolved binary, with methods~\ref{item:fit}~\& \ref{item:recfit} providing unbiased estimates and similar uncertainties. If, for some other application, the model $\vmod_1$ obtained with method~\ref{item:fit} were to differ significantly from the starting guess $\vmod_0$ (method~\ref{item:nocorr}), it would certainly make sense to examine whether recursive fitting (method~\ref{item:recfit}) is needed. However, while it converged for the smooth models I tested, I have not proven that it will necessarily do so, in particular for less smooth models that may require it.

\section{Conclusion}
\label{sec:conclusion}

The standard covariance propagation, using the measurement values in in the the calculation, can result in a bias in the model parameters of a least-squares fit taking correlations into account. It will occur as soon as the error bars and covariances depend on the measured values, in particular when a normalisation factor, such as the instrumental transfer function of an interferometer, is obtained experimentally.  Some bias will even occur without correlations, but the effect is strongest when a large set of correlated data is modelled.  This is precisely the case in optical and infrared long-baseline interferometry, where the calibration of spectrally dispersed fringes easily yields $10^2$ to $10^3$ correlated data points.

While solutions exist that are either numerically expensive or require some care to be implemented \citep{BUR11,BEC12,NIS14}, I have shown with a simple example that there is an easy and cheap way to solve the issue. First an uncorrelated fit is performed to estimate the true values corresponding to the data.  Secondly, these estimates are used to determine the covariance matrix by error propagation. At last, this covariance matrix is used to perform a least squares model fit.

Alternatively, it is possible to obtain an (almost) unbiased estimate for the model parameters by ignoring correlations altogether, with the cost of a larger imprecision, under-estimated uncertainties, and a biased chi square. It is, at the moment, the approach taken in the vast majority of published studies in optical interferometry, as data processing pipelines of most instrument do not determine covariances. To my knowledge, \citet{LAC19} is the only work where Peelle's Pertinent Puzzle has been explicitly taken care of in optical interferometry.

\section*{Acknowledgements}
This work has made use of the Smithsonian/NASA Astrophysics Data System (ADS).  I thank the anonymous referee for reading the paper carefully and providing constructive remarks, many of which have resulted in changes to the manuscript. 

\appendix

\section{Analytical derivation}

To shorten summations in the derivation, I introduce:
\begin{align*}
    m_l    &= \sum_i \frac{\data_i^l}{\dev_i^2}, \quad\text{(moments)}\\
    d_{lk} &= \sum_{i \ne j} \frac{(\data_i^k-\data_j^k)^l}{\dev_i^2\dev_j^2}, \quad\text{(moments of differences)},\\
    e_l    &= \sum_i \error_i^l, \quad\text{(moments of errors)}\\
    f_{lk} &= \sum_{i, j} (\error_i^k - \error_j^k)^l \quad\text{(moment of error differences)}\\
\intertext{and note that the differences can be developed as}
    d_{21} &= 2m_0m_2 - 2m_1^2,\\
    f_{21} &= 2n e_2 -  2e_1^2,\\
    d_{41} &= 2m_0m_4 - 8m_1m_3 + 6m_2^2,\\
    d_{22} &= 2m_0m_4 - 2m_2^2.
\end{align*}

\subsection{Equation \ref{eq:mu}}
\label{ap:mu}

I rewrite Eq.~(\ref{eq:invcov}) as 
\begin{equation*}
    \{ {\ppp\vcov}^{-1} \}_{ij} = \frac{\delta_{ij}}{\dev_i^2}
         - \frac{\corr\reldev^2\data_i\data_j}
                {\dev_i^2\dev_j^2(1 + \corr\reldev^2 m_2)}.
\end{equation*}
I then proceed from Eq.~(\ref{eq:pppmod}):
\begin{align*}
    \ppp{\mod} &= \frac{\sum\limits_{i,j} \invcov{ij} \data_i} 
                       {\sum\limits_{i,j} \invcov{ij} }\\
               &= \frac{\sum\limits_i \frac{\data_i}{\dev_i^2}
                     -  \frac{\corr\reldev^2}{1+ \corr\reldev^2m_2} 
                \sum\limits_{i,j} \frac{\data_i^2\data_j}{\dev_i^2\dev_j^2}}
                  {\sum\limits_i \frac 1{\dev_i^2}
                     -  \frac{\corr\reldev^2}{1+ \corr\reldev^2m_2} 
                \sum\limits_{i,j} \frac{\data_i\data_j}{\dev_i^2\dev_j^2}}\\
\intertext{so that, by separating summations in $i$ and $j$,}
               &= \frac
                   {m_1-\frac{\corr\reldev^2m_1m_2}{1 + \corr\reldev^2 m_2}}
                   {m_0-\frac{\corr\reldev^2 m_1^2}{1 + \corr\reldev^2 m_2}}\\
               &= \frac{m_1}{m_0 + \corr\reldev^2(m_0m_2 - m_1^2)}\\
               &= \frac{m_1}{m_0} \left( 1 + \corr\reldev^2\frac{d_{21}}{2m_0} \right)^{-1}.
\end{align*}
Equation~(\ref{eq:mu}) is obtained by noting that summation over $i < j$ is
half of that over $i, j$.

\subsection{Equation \ref{eq:mutaylor}}
\label{ap:mutaylor}

Since $d_{12}$ is expressed in terms of $(\data_i - \data_j)^2 = (\error_i - \error_j)^2$, the previous equation can be simplified in the second order as
\begin{align*}
    \ppp{\mod} &\approx \frac{m_1}{m_0} 
                - \corr\reldev^2
                    \left.  \frac{m_1}{2m_0^2} \right|_{\verror = \vec 0} 
                    \left. d_{21}\right|_{\dev_i = \devmean} 
                + o(||\verror||^2)\\
               &= \frac{m_1}{m_0} - \frac{\corr\reldev^2}{2n\stadev^2\datamean}  f_{21} + o(||\verror||^2).
\end{align*}

I now determine the Taylor development for
\begin{align*}
    m_l &= \frac{\datamean^{l-2}}{\stadev^2} \sum_i (1 + \error_i/\datamean)^{l-2}\\
    m_l &\approx \frac{n\datamean^{l-2}}{\stadev^2} \left[ 
                1 + \frac{l-2}{n\datamean}\sum_i \error_i 
               + \frac{(l-2)(l-3)}{2n\datamean^2}\sum_i \error_i^2\right]\\ 
    m_l &\approx \frac{n\datamean^{l-2}}{\stadev^2}\left[
            1 + \frac{l-2}{n\datamean} e_1 
                + \frac{(l-2)(l-3)}{2n^2\datamean} e_2 \right],
\end{align*}

The Taylor developments for $m_0$ and $m_1$ yield
\begin{align*}
   \frac{m_1}{m_0} &\approx \datamean
    \left[ 1 + \frac{e_1}{n\datamean} + \frac{2e_1^2 - 2n e_2}{n^2\datamean^2}    \right],\\ 
   \frac{m_1}{m_0} &\approx \datamean
    \left[ 1 + \frac{e_1}{n\datamean} - \frac{f_{21}}{n^2\datamean^2}    \right], 
\end{align*}
so that
\begin{equation*}
    \ppp{\mod} = \datamean + \frac{e_1}{n} 
                    - \frac1\datamean
                      \left( \frac 1{2n}\frac{\corr\reldev^2}{\stadev^2} + \frac 1{n^2} \right) 
                      f_{21}.
\end{equation*}

\subsection{PPP with photon detection}
\label{ap:photon}

I model $n$ photon measurements $N_i$ of expected value $\mean N$ with noise uncertainty $y_i = \sqrt{N_i}$ showing correlation $\corr$\footnote{Correlation in the photon noise is a quantum effect detected in particular experimental setting such as coupled lasers \citep[e.g.][]{MAY03}.  In astronomy, intensity interferometry makes use of these correlations.}  under the assumption of Gaussian errors ($\mean N \gg 1$). The statistical component of the uncertainty is given by $\dev_i^2 = (1-\corr)N_i$. The correlation matrix is
\begin{align*}
    \ppp{\cov}_{ij} &= \dev_i^2\delta_{ij} + \corr y_i y_j\\
\intertext{and its (Woodbury) inverse}
    \invcov{ij}    &= \frac{\delta_{ij}}{\dev_i^2} 
            - \frac{\corr y_iy_j}{\dev_i^2\dev_j^2(1 + \corr m'_2)},
\end{align*}
with the moments $m'_l$, $d'_{lk}$, etc. defined with respect to $y_i$, while
$m_l$, $d_{lk}$, etc. are defined with respect to $N_i$.

The least squares estimate for the number of photons is given by
\begin{align*}
    \ppp{\mod} &= \frac { \tr\vsens{\ppp\vcov}^{-1} \vec N }
                        { \tr\vsens{\ppp\vcov}^{-1} \vsens }\\
\intertext{and, by using $N_i = y_i^2$,}
    \ppp{\mod} &= \frac{\sum\limits_{i,j} \invcov{ij} y_i^2} 
                       {\sum\limits_{i,j} \invcov{ij} }\\
    \ppp{\mod} &= \frac{\sum\limits_i \frac{y_i^2}{\dev_i^2}
                     -  \frac{\corr}{1+ \corr m'_2} 
                \sum\limits_{i,j} \frac{y_i^3y_j}{\dev_i^2\dev_j^2}}
                  {\sum\limits_i \frac 1{\dev_i^2}
                     -  \frac{\corr}{1+ \corr m'_2} 
                \sum\limits_{i,j} \frac{y_iy_j}{\dev_i^2\dev_j^2}}\\
\intertext{so that, separating summations along $i$ and $j$,}
    \ppp{\mod} &= \frac{m'_2 - \frac{\corr m'_1m'_3}{1 + \corr m'_2}}
                       {m'_0 - \frac{\corr {m'_1}^2}{1 + \corr m'_2}},\\
    \ppp{\mod} &= \frac{m'_2 + \corr({m'_2}^2 - m'_1m_3)}
                       {m'_0 + \corr(m'_0m'_2 - {m'_1}^2)},\\
    \ppp{\mod} &= \frac{m'_2}{m'_0} 
                    \left(1 - \corr\frac{d'_{22} - d'_{41}}{8m'_2}\right)
                    \left(1 + \corr\frac{d'_{21}}         {2m'_0}\right)^{-1}\\
\intertext{or, more explicitly,}
    \ppp{\mod} &= \frac n{\sum\limits_i N_i^{-1}}
                  \frac{ 
                     1 - \frac{\corr}{2n(1-\corr)} 
                     \sum\limits_{i,j} 
                        \frac{(\sqrt{N_i}-\sqrt{N_j})^2}{\sqrt{N_i N_j}} 
                  }{ 1  + \frac{\corr}{2(1-\corr)\sum\limits_i N_i^{-1}} 
                     \sum\limits_{i,j} 
                        \frac{(\sqrt{N_i} - \sqrt{N_j})^2}{N_iN_j}}
                                  . 
\end{align*}

In the second order in $\verror$, it can be simplified to
\begin{align*}
   \ppp{\mod} &= \frac n{\sum\limits_i N_i^{-1}}
                    \left( 1 
            - \frac{\corr}{2n\mean{N}(1-\corr)} 
                \sum_{i,j} (\sqrt{N_i}-\sqrt{N_j})^2 \right)
\\
\intertext{and, by noting that $\sqrt{N_i} - \sqrt{N_j} = (N_i - Nj) / (\sqrt{N_i} + \sqrt{N_j}) \approx (\error_i - \error_j) / (2\sqrt{\mean{N}})$,}
   \ppp{\mod} &= \frac n{\sum\limits_i N_i^{-1}} \left(
                1 - \frac{\corr}{4n\mean{N}^2(1-\corr)} 
                \sum_{i,j} (\error_i - \error_j)^2 
            \right).
\end{align*}

The leading factor can be approximated in the second order using the Taylor series:
\begin{align*}
    \sum_i N_i^{-1} \approx \frac{n}{\mean{N}}
                           - \frac{e_1}{\mean{N}^2} 
                           + \frac{e_2}{\mean{N}^3}\\
\intertext{so that}
    \frac{n}{\sum\limits_i N_i^{-1}} \approx
            \mean{N} + \frac{e_1}{n} + \frac{e_1^2-ne_2}{n^2\mean{N}},\\
    \frac{n}{\sum\limits_i N_i^{-1}} \approx
            \mean{N} + \frac{e_1}{n} - \frac{f_{21}}{2n^2\mean{N}}.
\end{align*}

Finally,
\begin{align*}
    \ppp{\mod} &\approx \mean N  + \frac{\sum\limits_i \error_i}{n}
            - \left(1 + \frac{n\corr}{2(1-\corr)} \right)
                    \frac{\sum\limits_{i,j} (\error_i - \error_j) ^2}{2Nn^2}.
\end{align*}

With $\expect{(\error_i-\error_j)^2} = 2(1-\corr)\mean N$ and $\expect{\error_i} = 0$, the bias of the best fit estimate for the average number of photons is:
\begin{align*}
    \expect{\ppp{\mod}} &\approx \mean N - \left(1 - \frac 1n\right) \left( 
                    (1-\corr) + \frac{n\corr}{2}\right)\\
\intertext{or, with the relative statistical and systematic uncertainties $\stadev^2 = (1-\corr)/\mean{N}$ and $\sysdev^2 = \corr/\mean{N}$,}
    \expect{\ppp{\mod}} &\approx \mean N \left[ 1 - \left(1 - \frac 1n\right)
                \left(\stadev^2 + \frac n2 \sysdev^2\right)\right].
\end{align*}
The bias from Peelle's pertinent puzzle is exactly half of that determined for normalisation errors in the main part of the paper.  It shows that the effect does not necessarily arise from a normalisation.

\bibliographystyle{pasa-mnras}
\bibliography{article}

\begin{thebibliography}{}
\makeatletter
\relax
\def\mn@urlcharsother{\let\do\@makeother \do\$\do\&\do\#\do\^\do\_\do\%\do\~}
\definecolor{darkblue}{rgb}{0,0,0.597656}
\def\mndoi{\begingroup\mn@urlcharsother \@ifnextchar [ {\mndoi@} {\mndoi@[]}}
\def\mndoi@[#1]#2{\def\@tempa{#1}\ifx\@tempa\@empty \href
  {http://dx.doi.org/#2} {\textcolor{darkblue}{doi:#2}}\else \href
  {http://dx.doi.org/#2} {\textcolor{darkblue}{#1}}\fi \endgroup}
\def\mn@eprint#1#2{\mn@eprint@#1:#2::\@nil}
\def\mn@eprint@arXiv#1{\href {http://arxiv.org/abs/#1} {{\tt arXiv:#1}}}
\def\mn@eprint@dblp#1{\href {http://dblp.uni-trier.de/rec/bibtex/#1.xml}
  {dblp:#1}}
\def\mn@eprint@#1:#2:#3:#4\@nil{\def\@tempa {#1}\def\@tempb {#2}\def\@tempc
  {#3}\ifx \@tempc \@empty \let \@tempc \@tempb \let \@tempb \@tempa \fi \ifx
  \@tempb \@empty \def\@tempb {arXiv}\fi \@ifundefined
  {mn@eprint@\@tempb}{\@tempb:\@tempc}{\expandafter \expandafter \csname
  mn@eprint@\@tempb\endcsname \expandafter{\@tempc}}}

\bibitem[\protect\citeauthoryear{{Absil} et~al.,}{{Absil} et~al.}{2006}]{ABS06}
{Absil} O.,  et~al., 2006, \mndoi [\aap] {10.1051/0004-6361:20054522}, 452, 237

\bibitem[\protect\citeauthoryear{{Becker} et~al.,}{{Becker}
  et~al.}{2012}]{BEC12}
{Becker} B.,  et~al., 2012, \mndoi [Journal of Instrumentation]
  {10.1088/1748-0221/7/11/P11002}, 7, P11002

\bibitem[\protect\citeauthoryear{{Berger} et~al.,}{{Berger}
  et~al.}{2006}]{BER06}
{Berger} D.~H.,  et~al., 2006, \mndoi [\apj] {10.1086/503318}, 644, 475

\bibitem[\protect\citeauthoryear{{Burr}, {Kawano}, {Talou}, {Pan}  \&
  {Hengartner}}{{Burr} et~al.}{2011}]{BUR11}
{Burr} T.,  {Kawano} T.,  {Talou} P.,  {Pan} F.,   {Hengartner} N.,  2011,
  Algorithms, 4, 28

\bibitem[\protect\citeauthoryear{{D'Agostini}}{{D'Agostini}}{1994}]{DAG94}
{D'Agostini} G.,  1994, Nuclear Instruments and Methods in Physics Research A,
  346, 306

\bibitem[\protect\citeauthoryear{{ESO GRAVITY pipeline team}}{{ESO GRAVITY
  pipeline team}}{2020}]{GRAVITYpipe}
{ESO GRAVITY pipeline team} 2020, GRAVITY pipeline user manual Issue 1.4

\bibitem[\protect\citeauthoryear{{ESO MATISSE pipeline team}}{{ESO MATISSE
  pipeline team}}{2020}]{MATISSEpipe}
{ESO MATISSE pipeline team} 2020, MATISSE pipeline user manual Issue 1.5.1

\bibitem[\protect\citeauthoryear{{Eisenhauer} et~al.,}{{Eisenhauer}
  et~al.}{2011}]{GRAVITY}
{Eisenhauer} F.,  et~al., 2011, The Messenger, 143, 16

\bibitem[\protect\citeauthoryear{{Hummel} \& {Percheron}}{{Hummel} \&
  {Percheron}}{2006}]{MIDI}
{Hummel} C.~A.,  {Percheron} I.,  2006, in Society of Photo-Optical
  Instrumentation Engineers (SPIE) Conference Series. p. 62683X,
  \mndoi{10.1117/12.671337}

\bibitem[\protect\citeauthoryear{{Kammerer}, {M{\'e}rand}, {Ireland}  \&
  {Lacour}}{{Kammerer} et~al.}{2020}]{KAM20}
{Kammerer} J.,  {M{\'e}rand} A.,  {Ireland} M.~J.,   {Lacour} S.,  2020, \mndoi
  [\aap] {10.1051/0004-6361/202038563}, 644, A110

\bibitem[\protect\citeauthoryear{{Lachaume}}{{Lachaume}}{2003}]{LAC03}
{Lachaume} R.,  2003, \mndoi [\aap] {10.1051/0004-6361:20030072}, 400, 795

\bibitem[\protect\citeauthoryear{{Lachaume}, {Rabus}, {Jord{\'a}n}, {Brahm},
  {Boyajian}, {von Braun}  \& {Berger}}{{Lachaume} et~al.}{2019}]{LAC19}
{Lachaume} R.,  {Rabus} M.,  {Jord{\'a}n} A.,  {Brahm} R.,  {Boyajian} T.,
  {von Braun} K.,   {Berger} J.-P.,  2019, \mndoi [\mnras]
  {10.1093/mnras/stz114}, 484, 2656

\bibitem[\protect\citeauthoryear{{Lawson}}{{Lawson}}{2000}]{LAW00}
{Lawson} P.~R.,  ed. 2000, {Principles of Long Baseline Stellar Interferometry}

\bibitem[\protect\citeauthoryear{{Le Bouquin} et~al.,}{{Le Bouquin}
  et~al.}{2011}]{PIONIER}
{Le Bouquin} J.-B.,  et~al., 2011, \mndoi [\aap] {10.1051/0004-6361/201117586},
  535, A67

\bibitem[\protect\citeauthoryear{Mayer, Rana  \& Ram}{Mayer
  et~al.}{2003}]{MAY03}
Mayer P.~M.,  Rana F.,   Ram R.~J.,  2003, \mndoi [Applied Physics Letters]
  {10.1063/1.1539548}, 82, 689

\bibitem[\protect\citeauthoryear{{Millour}, {Valat}, {Petrov}  \&
  {Vannier}}{{Millour} et~al.}{2008}]{AMBER}
{Millour} F.,  {Valat} B.,  {Petrov} R.~G.,   {Vannier} M.,  2008, in Optical
  and Infrared Interferometry. p. 701349 (\mn@eprint {arXiv} {0807.0291}),
  \mndoi{10.1117/12.788707}

\bibitem[\protect\citeauthoryear{{Monnier}}{{Monnier}}{2007}]{MON07}
{Monnier} J.~D.,  2007, \mndoi [{New Astronomy Reviews}]
  {10.1016/j.newar.2007.06.006}, 51, 604

\bibitem[\protect\citeauthoryear{{Neudecker}, {Fr{\"u}hwirth}  \&
  {Leeb}}{{Neudecker} et~al.}{2012}]{NEU12}
{Neudecker} D.,  {Fr{\"u}hwirth} R.,   {Leeb} H.,  2012, Nuclear Science and
  Engineering, 170, 54

\bibitem[\protect\citeauthoryear{{Neudecker}, {Fr{\"u}hwirth}, {Kawano}  \&
  {Leeb}}{{Neudecker} et~al.}{2014}]{NEU14}
{Neudecker} D.,  {Fr{\"u}hwirth} R.,  {Kawano} T.,   {Leeb} H.,  2014, Nuclear
  Data Sheets, 118, 364

\bibitem[\protect\citeauthoryear{{Nisius}}{{Nisius}}{2014}]{NIS14}
{Nisius} R.,  2014, European Physical Journal C, 74, 3004

\bibitem[\protect\citeauthoryear{{Pauls}, {Young}, {Cotton}  \&
  {Monnier}}{{Pauls} et~al.}{2005}]{OIFITS1}
{Pauls} T.~A.,  {Young} J.~S.,  {Cotton} W.~D.,   {Monnier} J.~D.,  2005,
  \mndoi [\pasp] {10.1086/444523}, \href
  {http://adsabs.harvard.edu/abs/2005PASP..117.1255P} {117, 1255}

\bibitem[\protect\citeauthoryear{{Peelle}}{{Peelle}}{1987}]{PEE87}
{Peelle} R.~W.,  1987, Informal memorandum, Peelle’s Pertinent Puzzle.
Oak Ridge National Laboratory

\bibitem[\protect\citeauthoryear{{Perrin}}{{Perrin}}{2003}]{PER03}
{Perrin} G.,  2003, A\&A, 400, 1173

\bibitem[\protect\citeauthoryear{{Perrin}, {Ridgway}, {Coud{\'e} du Foresto},
  {Mennesson}, {Traub}  \& {Lacasse}}{{Perrin} et~al.}{2004}]{PER04}
{Perrin} G.,  {Ridgway} S.~T.,  {Coud{\'e} du Foresto} V.,  {Mennesson} B.,
  {Traub} W.~A.,   {Lacasse} M.~G.,  2004, \mndoi [\aap]
  {10.1051/0004-6361:20040052}, 418, 675

\bibitem[\protect\citeauthoryear{{Tallon-Bosc} et~al.,}{{Tallon-Bosc}
  et~al.}{2008}]{TAL08}
{Tallon-Bosc} I.,  et~al., 2008, in Optical and Infrared Interferometry. p.
  70131J, \mndoi{10.1117/12.788871}

\bibitem[\protect\citeauthoryear{{Tatulli} et~al.,}{{Tatulli}
  et~al.}{2007}]{TAT07}
{Tatulli} E.,  et~al., 2007, \mndoi [\aap] {10.1051/0004-6361:20064799}, 464,
  29

\bibitem[\protect\citeauthoryear{{Thi{\'e}baut}}{{Thi{\'e}baut}}{2008}]{THI08}
{Thi{\'e}baut} E.,  2008, in Optical and Infrared Interferometry. p. 70131I,
  \mndoi{10.1117/12.788822}

\makeatother
\end{thebibliography}

\end{document}